\documentclass[12pt]{iopart}
\usepackage{iopams,amssymb}
\usepackage{latexsym}
\usepackage{graphicx}
\newcommand{\ket}[1]{\left\vert#1\right\rangle}

\newcommand{\kebra}[2]{|#1\rangle\langle#2\vert}

\newcommand{\eq}{Eq.~}
\newcommand{\eqs}{Eqs.~}
\newcommand{\fig}{Fig.~}

\newcommand{\cf} {cf.~}
\newcommand{\ug} {\!=\!}

\newcommand{\piu} {\!+\!}
\newcommand{\meno} {\!-\!}

\newcommand{\rrefs} {Refs.~}

\newcommand{\eqref}[1]{(\ref{#1})}

\providecommand{\ket}[1]{|#1\rangle}

\providecommand{\kebra}[2]{\ket{#1}\!\bra{#2}}

\begin{document}
%

\title{Emergence of non-Markovianity in the emission process of an atom in a half-cavity}
\author{Tommaso Tufarelli{$^{1}$}, M. S. Kim{$^{1}$} and Francesco Ciccarello{$^{2}$},}

\address{{$^{1}$}{QOLS, Blackett Laboratory, Imperial College London, SW7 2BW, UK}}
\address{{$^{2}$}{NEST, Istituto Nanoscienze-CNR and Dipartimento di Fisica e Chimica, Universit$\grave{a}$  degli Studi di Palermo, via Archirafi 36, I-90123 Palermo, Italy}}
\begin{abstract}
We study quantum non-Markovianity in the early stage of the emission process of a two-level atom coupled to a semi-infinite waveguide, where the waveguide termination behaves as a perfect mirror. Specifically, we restrict to the analysis of the process for times shorter than twice the time delay $t_d$, where $t_d$ is the duration of a round trip along the atom-mirror path. We show the emergence of a threshold in the parameters space separating the Markovian and non-Markovian regions.
\end{abstract}

\maketitle
\section{Introduction}

Quantum non-Markovianity (NM) is currently the focus of a widespread interest \cite{petruccione,huelga,breuer}. Until recently such concept -- which is not as clear in the quantum realm as in classical physics -- has been associated with any open dynamics which is not governed by the well-known Kossakowski-Lindblad master equation \cite{petruccione}. Such paradigm has been considerably refined in the last few years through various investigations, which put forward quantitative measures of NM (see \cite{breuer} and references therein). Here, we make use of these developments in order to shed light on the occurrence of NM in the emission process of a quantum emitter (``atom") in a so-called half-cavity. Such model consists of a two-level system coupled to the electromagnetic field in the presence of a single perfect mirror. By hypothesis, the field is constrained to travel along a one-dimensional (1D) waveguide. One can think of a semi-infinite waveguide whose termination behaves as an effective perfect mirror. Such model was first studied in the 80s \cite{milonni} and, more recently, revisited and further developed in \rrefs\cite{zoller,tufa}. Studying NM in this system is especially interesting as the mirror clearly introduces a feedback mechanism, which can intuitively establish memory effects. One can however wonder whether or not these take place at any finite atom-mirror distance as well as if, and how, they are affected by interference phenomena occurring within the atom-mirror interspace. 

In general, the emission process under study is quite complex \cite{milonni, zoller,tufa}. The atomic excitation probability amplitude can be worked out exactly, although in the form of a function series featuring a number of terms that grows with time. Thereby a comprehensive analysis of NM, which will be the focus of an extensive publication \cite{tufa2}, cannot be attained in a fully analytical way. 
Nevertheless, there is numerical evidence that many distinctive qualitative traits of NM can be captured \cite{tufa2} even if one restricts the analysis to times that are shorter than $2t_d$, where $t_d$ is the duration of a round trip over the atom-mirror path. In such a case, the aforementioned series reduces to two terms only and an analytical description can be carried out. In this short work, we focus on such early stage of the time evolution ($t\!\le\!2t_d$) and demonstrate the existence of a {\it threshold} separating the Markovian and non-Markovian regions as a function of $\Gamma t_d$ ($\Gamma$ is the spontaneous emission rate in the no-mirror case) and $\phi$, the latter being the phase shift corresponding two twice the atom-mirror optical path. 

This paper is organized as follows. In Section \ref{model}, we present our model and review basic properties of the atomic emission process in a half-cavity. In Section \ref{criterion}, we describe the criterion we use to assess occurrence of NM. In Section \ref{threshold}, we show the existence of a threshold beyond which NM is exhibited. Finally, in  Section \ref{conclusions}, we give our conclusions.

\section{Model and main features of the emission process} \label{model} 
 We consider a two-level atom, with levels $\{\ket e,\ket g\}$ separated by the frequency $\omega_0$, interacting with a 1D continuum of bosonic modes $\hat a_k$, such that $[\hat a_k,\hat a_{k'}^\dagger]=\delta(k\!-\!k')$. Let the atom be placed at $x\ug x_0$, the $x$-axis being along the propagation direction of the field. We assume that the amplitude profile of the $k$-th mode is $\propto\sin{kx}$, which is equivalent to imposing the hard-wall boundary condition enforced by a perfect mirror lying at $x\ug0$. The Hamiltonian reads
\begin{equation}
\hat{H}\ug\omega_0\kebra{e}{e}\piu\!\int_0^{k_c}\!\!{\rm d}k\;\omega_k\hat a^\dagger_k\hat a_k\piu\int_0^{k_c}\!\!{\rm d}k\left(g_k\, \hat\sigma_+\hat a_k\piu\textrm{H.c.}\right),\label{H}
\end{equation}
where $\hat\sigma_+\ug\hat\sigma_-^\dag\ug\kebra{e}{g}$, $\omega_k\ug\omega_0\piu\upsilon (k\!-\!k_0),\,g_k\ug\sqrt{{\Gamma\upsilon}/{\pi}}\sin{k x_0}$ and $\upsilon$ is the group velocity of a photon (we assume a linear photonic dispersion relation). As we work under the Rotating Wave Approximation, the specific value of the cut-off wave vector $k_c$ is unimportant and hence we can make the approximation $\int_{0}^{k_c}{\rm d}k\simeq\int_{-\infty}^{\infty}{\rm d}k$ \cite{gardiner}. In this paper, we study the evolution of an arbitrary atomic state when the field is initially in the vacuum state $|0\rangle$. Since the ground state evolves trivially ($|g\rangle|0\rangle$ is an eigenstate of $\hat H$), we can focus on the evolution of the atomic excited state $\ket e$ only. Thus, at a time $t$ owing to conservation of the total number of excitations the joint atom-field wave function can be written as
$\ket{\Psi(t)}\ug\varepsilon(t)\ket e\!\ket 0\piu\ket{g}\!\int\!{\rm d}k\; \varphi(k,t) \,a^\dagger_k\!\ket{0}$
, where $\varepsilon(t)$ is the atomic excitation probability amplitude and $\varphi(k,t)$ is the single-photon amplitude in $k$-space. By imposing the time-dependent Schr\"{o}dinger equation, and eliminating the field modes in terms of $\varepsilon(t)$, one ends up with the delay differential equation \cite{milonni,zoller,tufa}
\begin{equation}
\dot{\varepsilon}(t)=-\frac{\Gamma}{2}\varepsilon(t)+\frac{\Gamma}{2}e^{i\phi}\varepsilon(t- t_d)\Theta(t- t_d)\,,\label{DE}
\end{equation}
where $ t_d=2x_0/\upsilon$ is the {\it time delay} , $\phi=2k_0x_0$ is a phase (optical path associated with $2 x_0$ and $k_0$) and $\Theta(t)$ is the Heaviside step function. For the initial condition $\varepsilon(0)\!=\!1$ the solution of \eq(\ref{DE}) reads
\begin{equation}
\varepsilon(t)=e^{-\frac{\Gamma t}{2}}\sum_n \frac{1}{n!}\left(\frac{\Gamma}{2}e^{ t_d/2}e^{i\phi}\right)^n\left(t-n t_d\right)^n\Theta(t-n t_d)\,\,.\label{eps}
\end{equation}
This suggests a natural splitting of the time-axis into intervals of duration $t_d$. In the first of such intervals the series in \eq(\ref{eps}) features only one term. When passing to the next interval, it acquires an extra term. 
It is easy to check that an arbitrary atomic initial state $\rho(0)$\small{$\!=\!\left(\begin{array}{cc} \rho_{gg}&\rho_{ge}\\
\rho_{eg}&\rho_{ee} \end{array}\right)$} at a later time $t$ evolves into
\begin{equation}
\rho(t)=\left(\begin{array}{cc}
(1-|\varepsilon(t)|^2)\rho_{gg}&\varepsilon^*(t)\rho_{ge}\\
\varepsilon(t)\rho_{eg}& |\varepsilon(t)|^2 \rho_{ee}
\end{array}\right).\label{matrice}
\end{equation}

\section{Criterion for assessing occurrence of non-Markovianity}\label{criterion}

To assess whether or not the emission dynamics exhibits NM, we use the following criterion based on the derivative of $|\varepsilon(t)|$. This is best expressed as
\begin{equation}
\frac{d}{dt}|\varepsilon(t)|\le0\,\,\,\,\,\,\,\,{\forall}\,t\label{cond}\,\,\,\,\,\,\,\,\Leftrightarrow\,\,\,\,\,\,\,\,{\rm the\,\, system\,\, is\,\, Markovian}.
\end{equation}
Hence, NM takes place if and only if $|\varepsilon(t)|$ (and in turn the atom's average energy) grows at a some stage of the evolution. The use of such criterion is justified by the fact that it can be proven \cite{laine} that an open dynamics of the form (\ref {matrice}) is divisible if and only if $d |\varepsilon(t)|/dt\!\le\!0$ at any time, where indivisibility is recognized as major trait of NM \cite{breuer}. Moreover, relevant and in general non-equivalent measures of NM, such as those in \rrefs \cite{laine, tore}, in the specific case of this dynamics vanish if and only condition (\ref{cond}) holds.

In the process under study, at times $t\!<\! t_d$ no memory effect can take place since a photon is not able to complete one round-trip between the atom and mirror. Indeed, (\ref{eps}) yields $|\varepsilon(t)|\!=\!e^{-\Gamma t/2}$ thus fulfilling (\ref{cond}) at any time $t\!\in\!$ [0,$ t_d$]. However, this may not be the case for $t\!>\! t_d$. To carry out our analysis, we first note that condition (\ref{cond}) is equivalent to the analogous condition ${d|\varepsilon(t)|^2}\!/{dt}\!\le\!0$. This, in the light of condition (\ref{cond}) and with the help of \eq(\ref{DE}) and its c.c., entails that in our case the system is Markovian iff
\begin{equation}
\frac{d}{dt}|\varepsilon(t)|^2=-\Gamma|\varepsilon(t)|^2+\Gamma{\rm Re}\left[e^{i \phi}\varepsilon(t- t_d)\varepsilon^*(t)\right]\le0\,\,\,\,\,\,\,\,\,\,\,\,\forall\,t\geq t_d.\label{cond2}
\end{equation}

\section{Non-Markovianity threshold in the time interval $[0,2t_d]$}\label{threshold}
 
Here, we focus on the conditions for occurrence of NM (in terms of the relevant parameters entering our open dynamics) in the first nontrivial time domain, namely $t\! \in\![ 0,2 t_d]$. Based on the previous section, in order to assess NM in the interval $[0,2t_d]$ it is enough to analyze the time domain $t\!\in\![t_d,2t_d]$. For such times, \eq(\ref{eps}) entails
\begin{equation}
\varepsilon(t)=e^{-\frac{\Gamma t}{2}} \left[1+\frac{\Gamma}{2}  e^{\frac{\Gamma  t_d }{2}+i \phi } (t- t_d )\right]\,.
\end{equation}
This and \eq(\ref{cond2}) then yield
\begin{equation}
\frac{{\rm d}|\varepsilon(t)|^2}{{\rm d}t}=-\frac{\Gamma e^{-\Gamma(x+ t_d)}}{4} \left(c_2 x^2+c_1 x+c_0\right)=-\frac{\Gamma e^{-\Gamma(x+ t_d)}}{4} p(x)\,,\label{px}
\end{equation}
where we have set $x\ug t- t_d$. Here, the coefficients specifying the second-degree polynomial $p(x)$ are given by
\begin{eqnarray}
c_2&=&\Gamma^2 e^{\Gamma t_d}\,\label{c2}\,,\\
c_1&=& -2 \Gamma e^{\frac{\Gamma  t_d }{2}} \left(e^{\frac{\Gamma  t_d }{2}}-2
   \cos \phi \right)\,\label{c1},\\
c_0&=& 4 \left(1-{e^{\frac{\Gamma  t_d}{2} }} \cos \phi \right)\,.\label{c0}
\end{eqnarray}
As the factor multiplying $p(x)$ in \eq(\ref{px}) is always negative, we see that the condition $d|\varepsilon(t)|^2/dt\!\le\!0$ for any $ t_d\!\le\!t\!\le2 t_d$ is equivalent to
\begin{equation}
p(x)\ge\!0\,\,\,\,\,\forall\,\,\,x\!\in\![0, t_d]\,.\label{cond3}
\end{equation}
If \eq(\ref{cond3}) holds then, in particular, it must be $p(0^+)\ug c_0\!\geq\!0$. Also, note that since $c_2\!>\!0$ for any $ t_d>0$ the parabola $p(x)$ is convex. In virtue of these considerations, we see that \eq(\ref{cond3}) is fulfilled iff one of the following conditions holds
\begin{itemize}
\item (i) \phantom{ii}$c_0\geq0$ and $\Delta<0$\,,
\item (ii) \phantom{i}$c_0\geq0$, $\Delta\geq0$ and $x_+\leq0$\,,
\item (iii) $c_0\geq0$, $\Delta\geq0$ and $x_-\geq t_d$\,,
\end{itemize}
where $\Delta\ug c_1^2\meno 4 c_0 c_2$ is the discriminant of the second-degree equation $p(x)\ug0$, while $x_\pm$ are the two real roots of $p(x)$ (when $\Delta\!\ge\!0$). Through elementary calculations
\begin{eqnarray}
\Delta=4 \Gamma^2 e^{\Gamma  t_d } \left(e^{\Gamma  t_d }-4\sin^2\phi\right)\,,\label{delta}\\
x_\pm(\phi, t_d)\ug\frac{e^{-\frac{\Gamma  t_d }{2}} }{\Gamma}\left(e^{\frac{\Gamma t_d}{2}}-2\cos\phi\pm\sqrt{e^{\Gamma t_d}-4\sin^2\phi}\,\right)\,.
\label{x1}
\end{eqnarray}
By comparing \eqs \eqref{c0} and \eqref{delta} it is easy to check that $\Delta<0$ implies $c_0\geq0$, so that (i) reduces to the simple inequality ${\rm e}^{\Gamma t_d/2}<2|\sin\phi|$.\\
Let us now address condition (ii). The requirements $c_0\geq0$ and $\Delta\geq0$ imply that it must be ${\rm e}^{-\Gamma  t_d/2}\geq\cos\phi$ and ${\rm e}^{\Gamma t_d/2}\geq2|\sin\phi|$. By imposing $x_+\leq0$, we find the further constraint $\cos\phi\geq\frac{1}{2}{\rm e}^{\Gamma t_d/2}$.\\
As for (iii), the previous considerations for (ii) hold unchanged except the condition on the root which is to be replaced with $x_-\geq t_d$. This results in the inequalities $\cos\phi\leq\frac12 {\rm e}^{\Gamma t_d/2}(1-\Gamma t_d)$ and $(1-\Gamma t_d)\cos{\phi}\leq\frac14{\rm e}^{\Gamma t_d/2}[(\Gamma t_d)^2-2\Gamma t_d]-{\rm e}^{-\Gamma t_d/2}$. 

\begin{figure}
\begin{center}
\includegraphics[width=.7\textwidth]{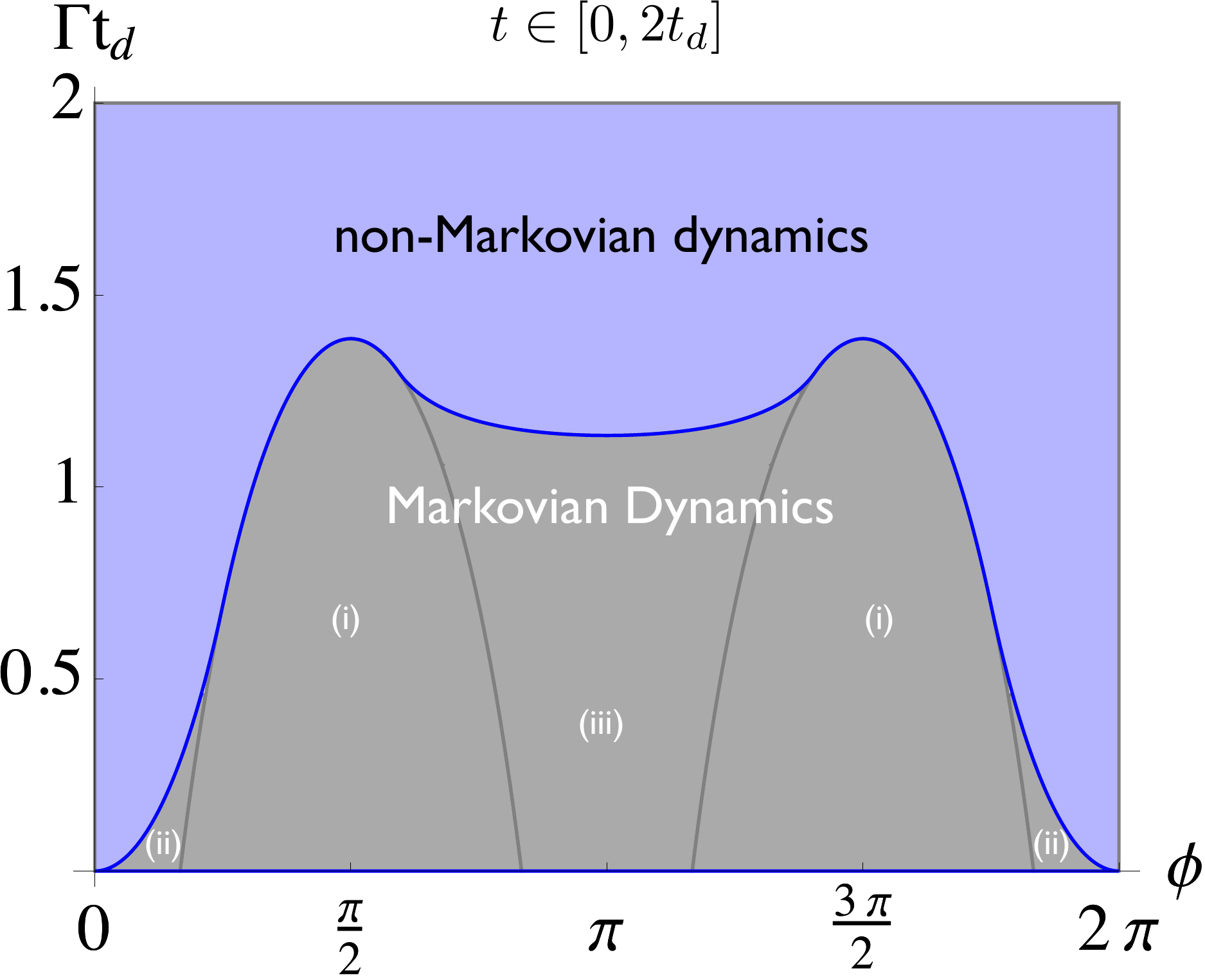}
\end{center}
\caption{Regions of Markovian and non-Markovian dynamics in the interval $t\!\in\![0,2 t_d]$. The threshold (blue thick line) separating the two regimes stems from the three domains (i), (ii) and (iii). \label{faiga}}
\end{figure}

In \fig\ref{faiga}, we display the three domains (i), (ii) and (iii) in the parameter space $\phi$-$\Gamma t_d$ (each region is intended as the domain within which the corresponding condition is fulfilled). NM occurs at points which do not lie within any of the three domains. 
By looking at \fig1, it is thus evident that the three domains in fact define a threshold in the parameter space separating the Markovian region from the non-Markovian one (thick line in \fig1). Remarkably, the threshold value of $\Gamma t_d$ beyond which NM is exhibited turns out to depend crucially on the phase $\phi$, hence witnessing occurrence of significant interference phenomena.

As observed in the Introduction, despite being limited to the early stage of the evolution, the above finding can be shown to capture some qualitative features arising from the analysis of the full evolution \cite{tufa2} (the latter is clearly demanded for a comprehensive and conclusive study of NM).  Intuitively, this is is because the time interval that we considered here is long enough for the system to exhibit crucial phenomena such as retardation, feedback and interference. Notwithstanding, some {\it general} conclusions (on the entire dynamics) based on our outcomes can be given. Indeed, a look at (\ref{cond}) should make clear that the condition that we found for occurrence of NM in terms of $\Gamma t_d$ at a set value of $\phi$ (\cf\fig1 and related discussion) is a {\it sufficient} one with respect to the entire time evolution. Hence, we conclude that the threshold arising from the full dynamics (if any) is upper bounded by the one in \fig1. In particular, two general properties can be established. First, for $\phi\ug2n\pi$ ($n$ is an integer number) the system is certainly non-Markovian {\it regardless of $\Gamma t_d$} (no threshold). Such feature is physically reasonable since it can be shown that these values of $\phi$ are the only ones for which an atom-photon bound state is formed \cite{tufa}. Second, NM surely occurs {\it independently of $\phi$} when $\Gamma t_d\!>\!2\ln2$. This is indeed the maximum value of $\Gamma t_d$ along the threshold, which occurs for $\phi\ug\pi/2$ (and even integer multiples of it, see \fig1).


\section{Conclusions} \label{conclusions}

We investigated the occurrence of NM for a two-level atom emitting in a 1D photonic waveguide in the presence of a perfect mirror. As a reliable criterion for assessing NM, we used the negativity of the time derivative of the atomic excitation probability. The study of NM has been limited to the early stage of the evolution, which allows to carry out most calculations analytically. We found the emergence of a threshold separating the Markovian and non-Markovian regions in the space of physical parameters. Based on this, we were able to highlight some properties concerning the full open dynamics.

\end{document}